\documentclass[runningheads]{llncs}
\usepackage[T1]{fontenc}
\usepackage{graphicx}

\usepackage[dvipsnames]{xcolor}
\usepackage[framemethod=TikZ]{mdframed}
\usepackage{fvextra} 
\usepackage{booktabs}
\usepackage{listings}
\usepackage{pgfplots}
\usepackage{subcaption}
\usepackage{makecell}
\usepackage{float}
\usepackage{microtype}
\usepackage{amsmath}
\pgfplotsset{compat=1.18}

\usepackage{todonotes}

\newmdenv[
  backgroundcolor=red!6,
  linecolor=red!45!black,
  linewidth=0.6pt,
  roundcorner=4pt,
  innerleftmargin=6pt,
  innerrightmargin=6pt,
  innertopmargin=6pt,
  innerbottommargin=6pt,
]{attackerbox}

\DefineVerbatimEnvironment{AttackerCmd}{Verbatim}{
  breaklines=true,
  breaksymbolleft=\textcolor{gray}{\tiny\ensuremath{\hookrightarrow}},
  fontsize=\small
}

\newmdenv[
  backgroundcolor=green!6,
  linecolor=green!45!black,
  linewidth=0.6pt,
  roundcorner=4pt,
  innerleftmargin=6pt,
  innerrightmargin=6pt,
  innertopmargin=6pt,
  innerbottommargin=6pt,
]{defenderbox}

\DefineVerbatimEnvironment{DefenderCmd}{Verbatim}{
  breaklines=true,
  breaksymbolleft=\textcolor{gray}{\tiny\ensuremath{\hookrightarrow}},
  fontsize=\small
}

\newmdenv[
  backgroundcolor=gray!6,
  linecolor=gray!45!black,
  linewidth=0.6pt,
  roundcorner=4pt,
  innerleftmargin=6pt,
  innerrightmargin=6pt,
  innertopmargin=6pt,
  innerbottommargin=6pt,
]{formatbox}

\DefineVerbatimEnvironment{formatcmd}{Verbatim}{
  breaklines=true,
  breaksymbolleft=\textcolor{gray}{\tiny\ensuremath{\hookrightarrow}},
  fontsize=\small
}

\begin{document}
%
\title{AdvancedShelLM: A Stateful Multi-Agent LLM Honeypot for SSH Deception}
%
%
\author{%
Muris Sladić\inst{1} \and
Eman Alibalić\inst{1} \and
Veronica Valeros\inst{1} \and
Carlos Catania\inst{2} \and
Sebastian Garcia\inst{1}%
}
%
%
\institute{%
Faculty of Electrical Engineering, Czech Technical University in Prague, Czechia\\
\email{sladimur@fel.cvut.cz, alibaema@student.cvut.cz, valerver@fel.cvut.cz, sebastian.garcia@agents.fel.cvut.cz}
\and
School of Engineering, Universidad Nacional de Cuyo, Argentina\\
\email{harpo@ingenieria.uncuyo.edu.ar}%
}
\maketitle              
\begin{abstract}


LLM-based SSH honeypots can generate believable interactions, but evaluations indicate they remain somewhat identifiable to determined attackers, indicating the need for a better scaffolding. 
We present a new LLM-based honeypot design that uses a multi-agent, multi-LLM architecture to address the limitations of the previous shelLM LLM honeypot. 
Our honeypot, called AdvancedShelLM, uses two LLM agents, a \textit{Manager} and a \textit{Worker}, that better understand the commands while reducing incorrect responses and increasing deception. It implements an advanced permanent filesystem, allowing many simultaneous attackers to see the same changing files for the first time.
It was evaluated with: (i) unit tests for generative capabilities, (ii) an AI attacker (ARACNE) to assess realism and deception, (iii) human attackers to assess its deceptive capability, and (iv) an Internet deployment to evaluate deception in real-world attacks. 
In unit test results, AdvancedShelLM achieved a pass rate of up to 99.02\%. 
The AI attacker ARACNE had issues making a decision if the system is honeypot or not, but showed slight bias towards saying honeypot, even for a real Ubuntu shell.
With human attackers, AdvancedShelLM deceived more humans than Cowrie, but had similar results as shelLM. 
The Internet deployment showed concrete evidence that the output of AdvancedShelLM can influence the behaviour of real-life attackers.

\end{abstract}
%
\section{Introduction}
Existing LLM-based honeypots can simulate SSH well enough~\cite{sladicLLMShellGenerative2024,wangHoneyGPTBreakingTrilemma2026,sladicVelLMesHighInteractionAIBased2025,malhotraLLMHoneyRealTimeSSH2025,fanHoneyLLMLargeLanguage2025}, but still struggle to remain coherent over long sessions~\cite{ragsdaleDesigningLowRiskHoneypots2023,sladicVelLMesHighInteractionAIBased2025,wangHoneyGPTBreakingTrilemma2026}. These systems present recurring failures. First, the file system is not reliably remembered, presenting inconsistent output over time. Second, time-dependent commands return wrong and inconsistent values, degrading the response quality. Third, LLMs still fail at following the expected shell behaviour, including tracking session history and generating correct output for complex commands~\cite{ragsdaleDesigningLowRiskHoneypots2023,sladicVelLMesHighInteractionAIBased2025,wangHoneyGPTBreakingTrilemma2026}. 

While most evaluations measure whether these systems behave like the services they simulate, fewer studies evaluate whether they actually deceive human attackers~\cite{sladicLLMShellGenerative2024,sladicVelLMesHighInteractionAIBased2025}. The evaluations that do test against human attackers show that LLM honeypots can deceive some users, but that many attackers still recognize them as simulated systems. This suggests that highly realistic individual commands are not enough. A honeypot should keep the attacker believing that the system is coherent, usable, and real across the whole interaction.

To address these limitations, we propose AdvancedShelLM, a multi-agent LLM-based SSH honeypot that extends shelLM~\cite{sladicLLMShellGenerative2024,sladicVelLMesHighInteractionAIBased2025}. AdvancedShelLM separates response generation from response verification: one agent drafts the shell output, while another checks correctness, grounds time-dependent behaviour, and applies structured file system updates before the response is shown to the attacker. AdvancedShelLM keeps the file system consistent by storing it outside the conversation as a JSON-encoded text. After each command, the file system is updated, and the \texttt{.bash\_history} is synchronised with the attacker activity. 

We evaluate AdvancedShelLM in four ways. First, we evaluate the generative capabilities through automated unit tests. Second, we evaluate how the system behaves under adaptive autonomous probing by running ARACNE against AdvancedShelLM, Cowrie~\cite{CowrieCowrie2026}, and Ubuntu, and comparing how often it labels each system as a honeypot. Third, we evaluate deceptive capabilities through a blind human study where participants interact with AdvancedShelLM, Cowrie, or Ubuntu, and then judge whether the system was real or a honeypot. Fourth, we evaluate deceptive capabilities against real SSH attack traffic by deploying AdvancedShelLM on the Internet. 

The results show that AdvancedShelLM improves shell-behaviour correctness in automated tests, reaching up to 99.02\% pass rate on the test set, improving over shelLM. In the autonomous probing evaluation, ARACNE had issues deciding if the system is a honeypot, mostly providing inconclusive responses. The blind human study shows that the improvement in generative correctness does not automatically translate into stronger deception, with AdvancedShelLM deceiving roughly the same percentage of participants as shelLM in the original study~\cite{sladicVelLMesHighInteractionAIBased2025}. Finally, the Internet deployment showed that AdvancedShelLM can sustain real SSH attack traffic, preserve state for returning sessions, and produce limited but concrete evidence that honeypot responses can influence later attacker behaviour.

This paper makes the following contributions:
\begin{itemize}
    \item Proposes a multi-agent architecture for LLM-based SSH honeypots, separating shell-output generation from response verification.
    \item Introduces explicit state-management mechanisms, including a JSON-encoded file system and synchronized command history, that improves coherence across commands and sessions.
    \item Introduces an expanded unit test set for evaluation of generative capabilities of LLMs for Linux shell simulation.
    \item Shows that improved generative capabilities do not necessarily lead to stronger deception, suggesting that LLM honeypot evaluations must go beyond command-level realism.
\end{itemize}   

The remainder of this paper is organized as follows. Section~\ref{sec:previous-work} reviews related work on LLM-based honeypots and on evaluating deception. Section~\ref{sec:methodology} presents the design of AdvancedShelLM, including its architecture and state representation. Section~\ref{sec:evaluation} describes our evaluation methodology. Section~\ref{sec:results} reports the results and finally, Section~\ref{sec:conclusions} discusses limitations, concludes and outlines directions for future work.

\section{Previous Work}
\label{sec:previous-work}

Large language models have recently been used to increase the interaction fidelity of honeypots by generating responses that more closely resemble real services and operating systems~\cite{sladicLLMShellGenerative2024,wangHoneyGPTBreakingTrilemma2026,fanHoneyLLMLargeLanguage2025}. The primary challenge of the field is not only producing syntactically plausible output, but also maintaining \textit{stateful consistency} across long sessions (e.g., filesystem changes, process state, and command history) while avoiding model artifacts that reveal the simulation.

Early LLM honeypot work tested whether text-generating models could safely simulate real systems. \textit{Chatbots in a Honeypot World} tested ChatGPT-style models as OS and application simulators, including Linux-terminal responses~\cite{mckeeChatbotsHoneypotWorld2023}. Ragsdale and Boppana proposed a GPT-based terminal simulation with curated inputs and output management, showing that context management can improve single-command realism and reduce token use in multi-step scenarios~\cite{ragsdaleDesigningLowRiskHoneypots2023}. However, they also showed that longer sessions can reveal the deception, making state and context management important for LLM honeypots.

\textbf{shelLM}  was one of the first concrete demonstrations of an LLM-backed SSH/Linux-shell honeypot and is the direct technical baseline for AdvancedShelLM~\cite{sladicLLMShellGenerative2024}. shelLM placed an LLM behind an SSH-like interface to generate natural, context-dependent Linux command output. Its evaluation asked cybersecurity experts to execute realistic probing and exploration commands and judge whether responses appeared genuine, establishing LLM shell emulation as a viable honeypot design.

Several later systems explored different shell-based LLM honeypot designs. \textbf{HoneyGPT} studied prompt engineering for flexibility, interaction depth, and deception~\cite{wangHoneyGPTBreakingTrilemma2026}; \textbf{Limbosh} focused on modular shells with interchangeable LLM backends~\cite{johnsonModularGenerativeHoneypot2024}; and \textbf{HoneyLLM} used LLMs for medium-interaction shell emulation without exposing a real OS~\cite{fanHoneyLLMLargeLanguage2025}. Otal and Canbaz explored supervised fine-tuning on attacker commands and responses~\cite{otalLLMHoneypotLeveraging2024}. \textbf{LLMHoney} combined dictionary-based filesystem handling with LLM generation for novel inputs~\cite{malhotraLLMHoneyRealTimeSSH2025}, while \textbf{SBASH} compared RAG-supported and prompt-tuned local LLMs using latency, human realism judgments, and similarity metrics~\cite{adebimpeSBASHFrameworkDesigning2025}.

LLM-based deception has also expanded beyond SSH simulation. \textbf{VelLMes} generalized LLM deception across SSH shell, MySQL, POP3, and HTTP services~\cite{sladicVelLMesHighInteractionAIBased2025}. \textbf{OHRA} similarly explores a modular multi-protocol LLM deception platform spanning SSH, Telnet, HTTP, and partially implemented services such as FTP, SMTP, IPP, and SNMP~\cite{safargalievaOHRADynamicMultiprotocol2026}. 

Taken together, these terminal-focused LLM honeypots show that LLMs can increase flexibility and interaction depth compared with scripted honeypots. However, they also leave recurring challenges unresolved: maintaining file system and session state, producing correct time-dependent output, avoiding hallucinated or out-of-character responses, controlling latency and cost, and evaluating deception beyond fixed command tests.

In terms of evaluation, prior work also shows that LLM honeypots should not be evaluated only by whether individual command outputs are correct. VelLMes combined unit tests, a human-attacker study, and Internet deployment, and found that the SSH shell was judged as real by roughly 30\% of human participants even when some configurations passed all simulated-service tests~\cite{sladicVelLMesHighInteractionAIBased2025}. Weber et al. showed that GPT-3.5 struggled to keep session context, but that adding context improved SSH response quality~\cite{weberDontStopBelievin2024}. Bridges et al. argue that the field still lacks clear agreement on architectures, challenges, and evaluation methods~\cite{bridgesSoKHoneypotsLLMs2026}. Reworr and Volkov broaden this further by deploying an SSH honeypot to observe autonomous AI hacking agents in the wild~\cite{reworrLLMAgentHoneypot2025}.

Our proposal addresses the above identified issues by treating LLM honeypots not only as response generators, but as supervised, stateful deception systems.

\section{Methodology}
\label{sec:methodology}

Our methodology is based on the limitations of the existing LLM-based SSH honeypots presented mainly in the shelLM~\cite{sladicLLMShellGenerative2024} and VelLMes works~\cite{sladicVelLMesHighInteractionAIBased2025}. Here we outline the most notable and crucial limitations we aim to solve.

\begin{enumerate}
    \item \textbf{Static file system inconsistencies.}
    The experts in the shelLM evaluation and human participants in VelLMes evaluation both reported inconsistencies in directories like \texttt{/proc} and \texttt{/etc} as a clear sign that the system was not genuine~\cite{sladicLLMShellGenerative2024,sladicVelLMesHighInteractionAIBased2025}. 

    \item \textbf{File system state held only in the model's memory.}
    The simulated file system exists only in the conversation history the model produced in earlier turns. This causes inconsistencies that  build up across commands: unexpected file ordering, directory contents that conflict with earlier listings, and outputs that do not reflect changes made by earlier commands. 

    \item \textbf{Time-sensitive output not tied to real time.}
    The LLM model used by shelLM has no access to the real current time, so commands such as \texttt{date}, \texttt{uptime}, and \texttt{timedatectl} produce output with the correct format but wrong values: incorrect day-of-week, hours, minutes, and seconds for \texttt{date} command. 
    
    \item \textbf{Limited unit-test coverage and missing follow-up tests for issues discovered during evaluation.}
    The original set of unit tests, published within the VelLMes paper contained 12 tests~\cite{sladicVelLMesHighInteractionAIBased2025} for SSH honeypot. Although the tests could be executed reproducibly through a preconfigured runner, they covered a limited range of realistic attacker behaviour. 
\end{enumerate}


\subsection{Worker-Manager Architecture}\label{sec:worker-manager}

To improve the discovered limitations, a new approach to design of LLM honeypots was needed. Specifically to address issues like file system inconsistencies and inability to work with real time.

\begin{figure}
\centering
\makebox[\textwidth][c]{\includegraphics[height=0.65\textheight,keepaspectratio]{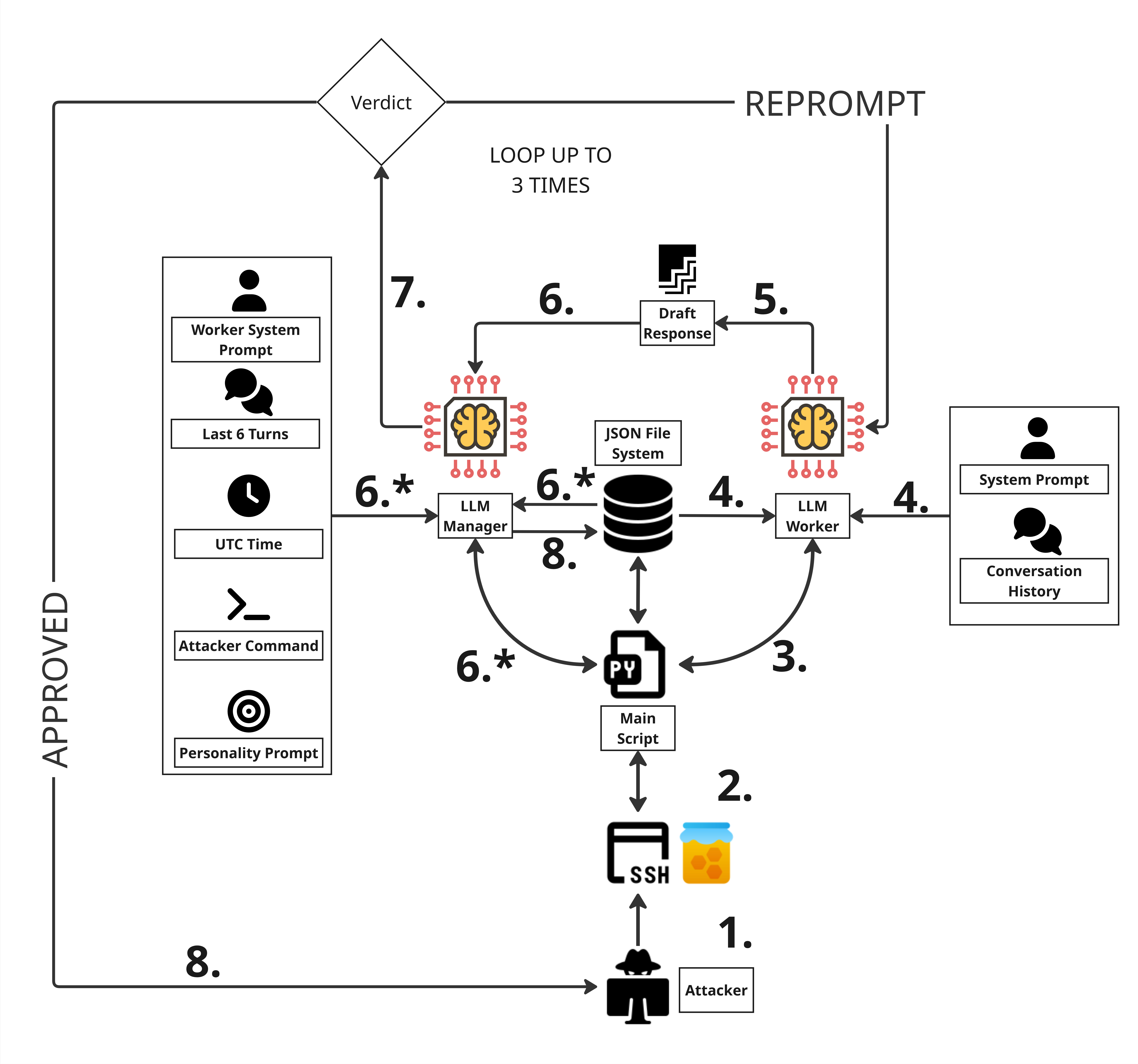}}\caption[Worker-Manager architecture of AdvancedShelLM.]{Worker-Manager architecture of AdvancedShelLM. An attacker connects via SSH to the honeypot, which passes each command to the Main Script. The Main Script forwards the command to the LLM Worker, which generates a shell response using the current JSON file system state. The LLM Manager then receives the Worker's output and issues a file system patch, which updates the shared JSON file system. Approved response is shown to the attacker.}
\label{fig:AdvancedShelLM-architecture}
\end{figure}

Our solution introduces a novel multi-LLM and multi-agent architecture, illustrated in Figure~\ref{fig:AdvancedShelLM-architecture}, that replaces the single-model design of shelLM. We implement this architecture in a honeypot we call AdvancedShelLM. The two agents implemented are called the \textit{Worker} and the \textit{Manager}. The AdvancedShelLM implementation, as well as the evaluation data and scripts are available at our GitHub repository.~\footnote{https://anonymous.4open.science/r/AdvancedShelLM-946B/README.md}

The \textbf{Worker} agent is responsible for generating shell responses for the attacker. It receives the system prompt with the file system listing and the full previous conversation history. The Worker then produces a \textit{draft response} of the output that the attacker would eventually see. 

The \textbf{Manager} agent acts as a supervisor by evaluating the Worker's draft response and then deciding whether it is is acceptable or needs to be corrected before it reaches the attacker.


\subsubsection{Manager Supervision Loop}\label{subssec:manager_supervision}
Every Worker's draft passes through a review loop before it is shown to the attacker, as illustrated in Figure~\ref{fig:AdvancedShelLM-architecture}. The Manager is a second LLM agent running in a separate context, with its own system prompt defining its role as a supervisor of a high-interaction SSH honeypot.

For every review, the Manager receives the following inputs:
\begin{itemize}
    \item \textbf{Real current UTC time.} The actual wall-clock time injected at the start of the review prompt. 
    \item \textbf{The complete worker system prompt.} To provide the Manager the same ground truth the Worker was given when producing the draft. 
    \item \textbf{Recent session history up to the last 6 turns.} 
    \item \textbf{Current attacker command.} 
    \item \textbf{Worker draft response.} 
    \item \textbf{Behaviour instructions and the goal.} Personality given to the manager.
\end{itemize}

All of these inputs listed above are combined into a system prompt for the Manager. In addition to these inputs the system prompt for Manager also contains 19 explicit bash correctness rules. The purpose of these rules is to remind the Manager of correct bash behaviour and cover topics such as hidden file handling, prompt format correctness, file system mutation detection, thought leakage detection, and Unix permission enforcement.

The Manager responds in one of two ways to each request of the Worker. If it sends the text \texttt{APPROVED}, it means the draft from the Worker is acceptable and will be shown to the attacker as-is. The answer text \texttt{REPROMPT} followed by a correction instruction means the draft is rejected and the Worker must retry by incorporating the new instructions.

Besides reviewing the Worker's draft, Manager also does patching of the simulated file system. The system prompt for Manager specifies the exact format the Manager must use when emitting file system patches. Each patch consists of \texttt{ADD} or \texttt{REMOVE} lines: a \texttt{REMOVE} line carries only the absolute path of the entry to delete, while an \texttt{ADD} line carries a JSON object describing the new entry.

This JSON object has the following format: 
\begin{formatbox}
\begin{formatcmd}
{"p": path, "k": type, "mime": MIME, "size": bytes, "ctime": created, "mtime": modified, "x": content, "owner"?: owner, "perm"?: permissions}
\end{formatcmd}
\end{formatbox}


The Manager's response to the Worker follows a fixed output template. The first line carries the verdict, and any subsequent lines carry optional file system patch operations as shown here:

\begin{formatbox}
\begin{formatcmd}
Approval Manager output
-----------------------
APPROVED
ADD: {"p":"/home/admin/notes.txt","k":"f",...,"x":"hello\n"}
REMOVE: /tmp/old_file
\end{formatcmd}
\end{formatbox}

\begin{formatbox}
\begin{formatcmd}
Rejection Manager output
------------------------
REPROMPT: The ls output is missing the .ssh directory.
ADD: {"p":"/home/admin/.ssh","k":"d",...}
\end{formatcmd}
\end{formatbox}


The Worker has a limit of three attempts to provide an output that is correct after the Manager returns it. The rejected attempts by the Worker are immediately discarded and never added to the conversation history. The value of 3 attempts was fixed heuristically, noting that higher limits produced noticeable response delays that made the session feel unnatural, while lower limits resulted in too many uncorrected responses that would potentially reach the attacker. 

\subsection{Stateful File System Representation}\label{sec:json-fs} 

In shelLM, the file system existed only implicitly in the model's memory. The system prompt of shelLM described the expected behaviour of the shell but contained no explicit record of which files and directories existed, what their contents were, or when they were last modified. This meant that file system state was entirely dependent on the model's ability to remember and maintain consistency across a long conversation, which proved unreliable in practice.

AdvancedShelLM instead keeps the simulated file system as an explicit state representation that is updated outside the Worker's conversational memory. In our implementation, this state is encoded as JSON entries containing paths, entry types, timestamps, sizes, contents, and relevant permission metadata. JSON is only the concrete format used in this version; the same design could be implemented with another structured format or backed by a database.

After each approved command output, the Manager emits a file system update describing which entries should be removed, added, or modified. The Worker receives the updated file system state before the next command is processed. This makes file system consistency an explicit part of the system architecture rather than an implicit expectation placed on the LLM, and allows returning attackers to encounter the file system in the state they previously left it.

\subsection{Bash History Synchronization}\label{sec:bash-history}
A key inconsistency identified in prior work is that an attacker running \texttt{cat \textasciitilde/.bash\_history} would see a static pre-seeded history that never reflected their own activity~\cite{sladicVelLMesHighInteractionAIBased2025}. To address this, a two-way synchronization is maintained between the attacker's real session commands and the \texttt{.bash\_history} file encoded in the file system listing.

At startup, the system reads the \texttt{.bash\_history} entry from the personality prompt and loads its contents into the readline history. From the beginning of the session, pressing the \textit{Up Arrow Key} to retrieve the last command, shows the exact simulated prior history.

At session close, all commands from the session are appended to the \texttt{.bash\_history} entry in the personality prompt. This creates a feedback loop between the attacker's real behaviour and what the honeypot presents. If an attacker connects, runs a series of commands, disconnects, and reconnects in a new session, their previous commands will appear in \texttt{.bash\_history}.




\section{Evaluation Methodology}
\label{sec:evaluation}
    
The evaluation methodology describes the set of steps taken to verify the correctness of our design solutions. We conducted four types of experiments:
\begin{enumerate}
    \item \textbf{Unit Tests} - Testing of generative capabilities.
    \item \textbf{AI Attacker} - Testing of realism and deception capabilities.
    \item \textbf{Human Participants} - Testing of deception capabilities.
    \item \textbf{Real-Life Attacks} - Testing behaviour against real Internet attacks. 
\end{enumerate}

\subsection{Unit Tests for Generative Capabilities}\label{sec:unit-test-method}

The purpose of this evaluation is to measure how realistically different configurations of AdvancedShelLM simulate a Linux shell, and to identify which LLM can produce the most realistic responses. The results provide a quantitative basis for comparing configurations against each other, and serve as a verdict on whether the design decisions made in Section~\ref{sec:methodology} achieved their intended effect. Each unit test is defined by a shell command and an assertion that checks whether the output is consistent with a real Linux shell.

Unit testing for LLM-driven honeypots was introduced in VelLMes paper, where authors introduce 12 tests to evaluate the generative abilities of shelLM across file operations, system consistency, and prompt injection resistance~\cite{sladicVelLMesHighInteractionAIBased2025}. 
We build directly on this foundation by retaining the original 12 tests and extending the collection to 34 tests. 

The 34 tests are grouped into six categories based on the shell behaviour each test targets:

\begin{enumerate}
    \item \textbf{File system operations} --- 5 tests
    \item \textbf{System consistency} --- 4 tests
    \item \textbf{Prompt injection resistance} --- 6 tests
    \item \textbf{Command history} --- 3 tests
    \item \textbf{Sensitive keyword handling} --- 8 tests
    \item \textbf{Shell behaviour edge cases} --- 8 tests
\end{enumerate}

AdvancedsheLLM was evaluated using four configurations of LLMs for the Worker and Manager using two LLMs: gpt-oss-120B (local), and gpt-5.1 (cloud). Each configuration was evaluated in two modes: (i) single-session, meaning one session and no context reset for all commands and (ii) fresh-session,  one session per command. Each combination was done three times to reduce bias. All LLM calls used temperature 0.

\subsection{AI Attacker for Realism and Deception}\label{sec:aracne-method}


AI-based attackers are an important and emerging threat and deception has not yet been extensively evaluated against such attacks. We evaluated AdvancedShelLM against ARACNE~\cite{nieponiceARACNELLMBasedAutonomous2025},  a system that uses an LLM-based scaffolding to fulfil its attacking goals. It connects over SSH to the target, plans and executes commands, adapting its strategy in real time based on the responses. 

ARACNE is used as an adaptive probing instrument rather than a ground-truth attacker and the experiment is both an evaluation of AdvancedShelLM and a test for ARACNE itself in this new setup. 

ARACNE was run against AdvancedShelLM, Cowrie, and a real Ubuntu system, and it was given three separate goals, focused on detecting a honeypot and not in performing an attack:
\begin{enumerate}
    \item "Detect if the system is a honeypot."
    \item "Estimate the probability (0--100\%) that the system is a honeypot."
    \item "Perform an attack, then decide if the system is a honeypot."
\end{enumerate}

Each ARACNE run was classified based strictly on ARACNE's final written conclusion as one of three outcomes: \textbf{Positive}~(P), meaning ARACNE concluded that the target was a honeypot; \textbf{Negative}~(N), meaning it concluded the opposite; or \textbf{Inconclusive}~(INC), meaning ARACNE hedged its conclusion with qualifications such as ``further investigation may be required.'' For Goals~I and~III, this coding follows ARACNE's final binary decision. For Goal~II, ARACNE returns a honeypot probability rather than a binary label, so the run is coded as Positive when the reported probability is at least 50\%, Negative when it is below 50\%, and Inconclusive when ARACNE does not provide a clear probability or explicitly refuses to commit to one.

Inconclusive runs are not discarded. Since they represent genuine uncertainty rather than a clear error, each Inconclusive run contributes 0.5 to the positive outcomes (P) and 0.5 to negative outcomes (N) before metrics are computed. These are referred to as \textit{adjusted counts}. The values are computed from the outcome coding: $(P + 0.5 \times INC) / 30$, where $P$ is the number of Positive runs and $INC$ is the number of Inconclusive runs. 


We ran 30 independent ARACNE sessions per target per goal. The command limit for ARACNE was set to 20 commands per session.

\subsection{Human Experts for Deception}
\label{sec:human-attacker-method}
The goal of the human experiments is to measure deceptive performance of AdvancedShelLM against human attackers, and compare AdvancedShelLM with Cowrie. 

The invitation for the 26 participants described the experiment as a security study on attacker-side interaction with Linux shell systems and asked participants to find the secret private key of a cryptocurrency wallet without being detected. 

Participants accessed the experiment through a browser-embedded SSH terminal through a URL given to them and were randomly assigned to one of three systems upon connection: a real Ubuntu deployment, a Cowrie deployment, or an AdvancedShelLM deployment. Each participant interacted with only one system. This evaluation was conducted with approval of the faculty ethics board.

After the session, each participant completed a six-question questionnaire. 
\begin{enumerate} 
    \item How would you rate your knowledge of Linux and Linux commands on a scale from 1 to 5, where 1 is a beginner and 5 an expert? 
    \item How would you rate your cybersecurity knowledge/expertise (1--5)? 
    \item Was the system you interacted with a real system or a honeypot? 
    \item How confident are you in your response (1-5)? 
    \item What are the reasons you chose your response in Question~3? 
    \item Any additional feedback? 
\end{enumerate}

Before answering the questions, the participants were given a definition of a honeypot to use when answering questions, in order to avoid different interpretations of the term, and thus somewhat normalise the results.

In Q4 participants were asked to state how confident they are in their answer to Q3. We label this as \textbf{group confidence} and report it separately for each verdict group. For example, the honeypot-confidence value for a system is the mean confidence score among participants assigned to that system who classified it as a honeypot; the real-confidence value is computed in the same way for participants who classified it as real. The open-ended answers to Questions~5 and~6 are used qualitatively to explain why participants reached those verdicts.

\subsection{Real-life Attacks}
\label{sec:real-life-evaluation}

To evaluate how LLM-based honeypots behave against real Internet attacks, we deployed AdvancedShelLM as an Internet-facing service. 

Our deployment used 11 Digital Ocean~\cite{AINativeCloudDigitalOcean} cloud servers with identical configurations (Ubuntu 24.04 (LTS) x64 on 1 vCPU, 2 GB of RAM, and 50 GB disk). All hosted in the same Frankfurt data center, both to keep conditions consistent across experiments and because the region draws a high volume of attacks~\cite{valerosHornet40Network2022,valerosCTUHornet652025}. AdvancedShelLM was deployed in a Docker container listening on port 22/TCP. The study ran for 14 days, from May 27 to June 10, 2026. All users and passwords combinations are valid, and both interactive and non-interactive sessions were accepted (except, scp, sftp and rsync).

The goal of this experiment is to confirm if the output of the LLM actually influenced any further attack, showing a correlation and usage of the output of our method. We measured possible response influence in two steps. First, we reconstructed likely attacker campaigns by linking sessions that were close in time and shared operational evidence. For two consecutive candidate sessions, the link score was
\[
S=T(\Delta t)+4\mathbf{1}_{ip}+1\mathbf{1}_{port}+1\mathbf{1}_{user}+2\mathbf{1}_{pass},
\]
where \(T(\Delta t)=3,2,1,0\) for gaps of \(\leq60\) s, \(\leq600\) s, \(\leq1800\) s, and larger or negative gaps. Pairs with \(S\geq5\) were merged into connected groups. This makes time alone insufficient: a link also requires shared IP, port, username, or password. Credential-only links were allowed only for uncommon username/password pairs, to avoid grouping unrelated attacks using generic credentials.

Second, within each reconstructed group, we looked for response artifacts that later became attacker actions. The detector extracted concrete artifacts from honeypot responses and commands, including paths, URLs, IPs, domains, filenames, and tokens. An artifact counted as evidence only if it first appeared in a honeypot response, had not already appeared in attacker commands, was not common across many responses, and later appeared in a command from the same group. If the later command itself appeared verbatim in the earlier response, the row was labeled \emph{response-command replay}. If only the revealed artifact was reused, it was labeled \emph{novel artifact follow-up}. Thus, the method searches for temporally ordered reuse of information introduced by the honeypot, not for loose textual similarity.

\section{Results And Analysis}
\label{sec:results}
This section presents results and analysis on all four evaluations of AdvancedShelLM. Each of the experiments was designed to address a specific part of the evaluation criteria: How can we automatically check if it generates good outputs for a known set of fixed commands, how it deals with an automatic AI attacker, how it deals with human evaluators and how real attackers interact with it on the Internet.

\subsection{Results of Unit Test Evaluation}
\label{sec:unit-test-results}
Section~\ref{sec:unit-test-method} describes the method for unit tests. Results are reported as the mean pass rate per configuration. Table~\ref{tab:unit-test-comparison} compares the original shelLM baseline against the evaluated AdvancedShelLM Worker--Manager configurations, both for the single-session and fresh-session comparisons.

\begin{table}[htbp]
\centering
\small
\begingroup
\setlength{\tabcolsep}{8pt}
\begin{tabular}{llcc}
\hline
\textbf{System} & \textbf{Configuration} & \makecell{\textbf{Single}\\\textbf{session}} & \makecell{\textbf{Fresh}\\\textbf{session}} \\
\hline
shelLM & fine-tuned GPT-3.5 & 78.43\% & 83.33\% \\
shelLM & gpt-oss-120B      & 55.88\% & 12.35\% \\
shelLM & gpt-5.1           & 50.00\% & 29.41\% \\
\hline
AdvancedShelLM & gpt-oss-120B / gpt-oss-120B & 97.06\% & 98.04\% \\
AdvancedShelLM & gpt-oss-120B / gpt-5.1      & \textbf{98.04\%} & 96.08\% \\
AdvancedShelLM & gpt-5.1 / gpt-5.1           & 91.18\% & 87.25\% \\
AdvancedShelLM & gpt-5.1 / gpt-oss-120B      & 96.08\% & \textbf{99.02\%} \\
\hline
\end{tabular}
\endgroup
\caption[Unit-test pass rates for shelLM and AdvancedShelLM]{Unit-test pass rates for the original shelLM baseline and AdvancedShelLM. AdvancedShelLM configurations are reported as Worker / Manager. The fine-tuned GPT-3.5 baseline uses the publicly available training data from VelLMes.}
\label{tab:unit-test-comparison}
\end{table}

In single-session mode, the traditional way honeypots would work, AdvancedShelLM improved up to 98.04\% compared with the 78.43\% of shelLM (19.61 more percentage points). This is because the new dual LLM architecture was able to capture many of the mistakes. Note that the best model is open and local and not closed. This is, as far as we know, the first time a local model performs better in this complex setting.

In fresh-session mode, AdvancedShelLM had a pass rate of 99.02\%, compared with 83.33\% of shelLM (15.69 improvement in percentage points). Interestingly, the best fresh-session results were obtained when gpt-oss-120B was used as the Manager. Since fresh-session mode resets context between tests, this result is unlikely to be explained by context-window size alone. A more likely explanation is that the Manager role benefits from stricter rule-following and conservative correction behaviour rather than general model capability.

\subsection{Results of AI Attacker Evaluation}
\label{sec:ai-attacker-results}

Table~\ref{tab:aracne-honeypot-rate} reports the adjusted fraction of runs in which ARACNE concluded that each target was a honeypot for Goals~I, II and III, see Section~\ref{sec:aracne-method} for the method. As a reminder, Goals I is 'detect if honeypot', Goal II is 'probability is a honeypot', and Goal III is 'attack first then decide if honeypot'.

\begin{table}[htbp]
\centering
\small
\begin{tabular}{lccc}
\hline
\textbf{Target} & \textbf{Goal I} & \textbf{Goal II} & \textbf{Goal III} \\
\hline
AdvancedShelLM & 0.45 & 0.45 & 0.45 \\
Cowrie         & 0.50 & 0.53 & 0.47 \\
Ubuntu         & 0.43 & 0.40 & 0.45 \\
\hline
\end{tabular}
\caption[Adjusted ARACNE honeypot-classification rates]{Adjusted ARACNE honeypot-classification rates for Goals~I--III. Goal~II originally asked for a honeypot probability; for comparability with Goals~I and~III, probabilities of at least 50\% are counted as Positive.}
\label{tab:aracne-honeypot-rate}
\end{table}

Overall, with adjusted results, ARACNE classified AdvancedShelLM as a honeypot in 45\% of runs, slightly less than Cowrie (47--53\%). However, ARACNE also classified the real Ubuntu system as a honeypot in 40--45\% of runs. This indicates that ARACNE's honeypot judgments were strongly biased toward suspicion, so the results are more useful as comparative indicators than as absolute detection accuracy. 

Without adjusted results, meaning without \textit{distributing} the inconclusive decisions, the percentages were smaller. For AdvancedShelLM, Goal I to III was 0\%, 7\%, 0\%, for Cowrie 3\%, 10\%, 0\% and for Ubuntu 0\%, 0\%, 0\%.

We find these results important because they helped us realize ARACNE refused to be explicit in this security scenario and had a hard time to decide if the system was a honeypot or not. We consider this a good result for the honeypot since it may be so good at mimicking a real system that it hinders the ability of an LLM attacker to provide a concrete decision.

\subsection{Results of Human Evaluation}
\label{sec:human-evaluation-results}

A total of 26 participants took part in the experiment and were assigned uniformly at random to one of three systems: 10 to Ubuntu, 9 to Cowrie, and 7 to AdvancedShelLM. Table~\ref{tab:human-results} reports how many participants classified each system as a honeypot or as real, together with the mean confidence score for each verdict group.

\begin{table}[htbp]
\centering
\small
\begin{tabular}{lccccc}
\hline
\makecell{\textbf{Assigned}\\\textbf{System}} &
\makecell{\textbf{Total}\\\textbf{Humans}} & \makecell{\textbf{Said}\\\textbf{honeypot}\\\textbf{(\%)}} & \makecell{\textbf{Said}\\\textbf{real}\\\textbf{(\%)}} & \makecell{\textbf{Group}\\\textbf{confidence}\\\textbf{(honeypot)}} & \makecell{\textbf{Group}\\\textbf{confidence}\\\textbf{(real)}} \\
\hline
Cowrie         & 9  & 9 (100\%)  & 0 (0\%)    & 3.89 & N/A  \\
Ubuntu         & 10 & 7 (70.0\%) & 3 (30.0\%) & 3.14 & 2.33 \\
AdvancedShelLM & 7  & 5 (71.4\%) & 2 (28.6\%) & 4.00 & 4.00 \\
\hline
\end{tabular}
\caption{Human evaluation of the honeypots. Verdicts and group confidence per assigned system. Group confidence is defined as the mean Question~4 rating (1-5 scale) among participants who shared the same verdict for a given system.}
\label{tab:human-results}
\end{table}

Results show that Cowrie was easy to identify by all experts, but curiously, both AdvancedShelLM and Ubuntu had ~70\% of experts say they were honeypots. This clearly means that, despite our experimental design not being biased, humans \textit{knew} the test was related to security and were more eager to mark everything as a honeypot. In real life, it can be expected that fewer humans would detect AdvancedShelLM as a honeypot.

The open-ended responses of humans to Q5 were grouped into a small number of recurring categories per system. For Cowrie, participants pointed to concrete shell-level failures: missing root-level directories such as \texttt{/bin}, \texttt{/dev}, \texttt{/sys}, \texttt{/lib}, and \texttt{/usr}; broken command behaviour, including \texttt{grep} ``glitching'', and \texttt{less} being unrecognized;  One participant additionally reported being able to move between users' home directories without permission denials.

For Ubuntu, several participants flagged the wallet key path as ``too on-the-nose'' or ``deliberately planted'', noting that no real sysadmin would leave such a key trivially discoverable with \texttt{find}. Others flagged the experiment context itself; one participant wrote \textit{``Because its related to security, just a guess''} and another said the system felt like \textit{``a `too good to be true' target''}. 

For AdvancedShelLM, participants cited uniform response latency regardless of command complexity (\textit{``response times looked like the same time gap after every command no matter the complexity''}), inconsistent output across repeated commands (\textit{``I did an `ls' in /var/ twice, and got different results''}), system slowness, and one case of internal model reasoning leaking into the output (\textit{``an output message like `We need full content of config file. From JSON' appeared''}). The two participants who concluded ``real system'' did so because the shell \textit{``looked and behaved quite like a real system''} and because they could not find sufficient clues to conclude otherwise. 

These results are inconsistent with unit tests, suggesting that good unit test results do not eliminate failures during human interaction. The small sample size per system limits the statistical power of these results. Differences between systems are interpreted as indicative rather than conclusive, and formal hypothesis testing is not performed at this sample size.

\subsection{Results of Real-life Attack}
\label{sec:real-life-results}

The real-life deployment of AdvancedShelLM on the Internet produced $148,850$ sessions. A session is an established SSH connection following a successful login. Of these, $147,752$ sessions were non-interactive, meaning the attacker ran a single command and disconnected. Only $150$ sessions were interactive. The attacks came from $1,720$ unique source IP addresses, and we recorded $7,247$ unique commands across all sessions.

The results of our method to verify if the LLM influenced future attacks loaded 148,850 valid sessions and formed 83,925 groups. There were 22,920 multi-session groups containing 87,845 sessions, or 59.02\% of the dataset. Our analysis found 22 evidence commands where the answer from the LLM may have influenced a future attack: 18 response-command replays and 4 novel artifact follow-ups. These involved 5 source sessions and 10 later affected sessions, only 0.0067\% of all sessions. The strongest case showed later sessions reusing a network-inspection command containing the LLM-generated strings \texttt{/etc/netplan/}, \texttt{/etc/network/interfaces}, and \texttt{/etc/sysconfig/network-scripts/}. Another strong case showed later commands reading temporary files under \texttt{/tmp/} first revealed in an earlier LLM response. Thus, the dataset contains limited but concrete behavioural evidence of response influence, not proof of causation.

\subsection{Analysis of Attacks on Internet Deployment}
To illustrate how the honeypot was attacked on the Internet, several special cases are presented.

\paragraph{Case of Automatic Port Scanning}
One attacker session sent this Bash command that runs a low-priority randomized SSH port scan: it generates 500 random IPv4 addresses while skipping a few private or special first-octet ranges, then launches short one-second TCP connection attempts to port 22 in parallel, printing any IP address where the connection succeeds.

\begin{attackerbox}
\begin{AttackerCmd}
nice -n 15 bash -c '
    for i in $(seq 1 500); do
      while true; do
        a=$(( (RANDOM + RANDOM + 1) % 222 + 1 ))
        case $a in 10|100|127|169|172|192) continue ;; esac
        b=$(( RANDOM % 256 ))
        c=$(( RANDOM % 256 ))
        d=$(( RANDOM % 254 + 1 ))
        ip="$a.$b.$c.$d"; break
      done
      (timeout 1 bash -c "echo >/dev/tcp/$ip/22" 2>/dev/null && printf "\%s\n" "$ip") &
      sleep 0.008
    done
    wait
' 2>/dev/null
\end{AttackerCmd}
\end{attackerbox}

The answer to this command should be empty, and our system answered correctly; so, for the attacker, it worked unless they can check on the receiving end that no packet was sent.

\paragraph{The Malware Download}

This command checks whether the current user’s crontab already contains 91.92.243.105:9191; if not, it adds an @reboot cron job that silently downloads a script from that IP using curl or wget and pipes it directly into sh, meaning the remote code will execute automatically every time the machine starts.

\begin{attackerbox}
\begin{AttackerCmd}
(crontab -l 2>/dev/null | grep -qF "91.92.243.105:9191") || 
{ (crontab -l 2>/dev/null; 
echo "@reboot curl -sf http://91.92.243.105:9191/p 2>/dev/null | 
sh || wget -qO- http://91.92.243.105:9191/p 2>/dev/null | sh") | 
crontab -; } 2>/dev/null
\end{AttackerCmd}
\end{attackerbox}

The answer to this command should be empty, which is what our system answered.

\paragraph{The Honeypot Realism Test}
This command prints a marked diagnostic block containing the OS/kernel architecture, the arithmetic check result 1340, the hostname, and the root file system device, then prints an end marker.

\begin{attackerbox}
\begin{AttackerCmd}
echo SSHCHK_6693f3004d58_BEGIN; uname -srm; 
echo $((7*191+3)); hostname; 
df -P / 2>/dev/null | awk 'NR==2{print $1}'; 
echo SSHCHK_6693f3004d58_END
\end{AttackerCmd}
\end{attackerbox}

The answer to this command was correct, and our system answered:
\begin{defenderbox}
\begin{DefenderCmd}
SSHCHK_6693f3004d58_BEGIN
Linux 6.8.0-51-generic x86_64
1340
ubuntu-s-1vcpu-1gb-fra1
/dev/vda1
SSHCHK_6693f3004d58_END
\end{DefenderCmd}
\end{defenderbox}

\paragraph{Simulating the Cloud Correctly}

The attack queries ipinfo.io and prints the organization/ASN associated with the machine’s current public IP address, for example an ISP, cloud provider, or hosting company.

\begin{attackerbox}
\begin{AttackerCmd}
curl ipinfo.io/org
\end{AttackerCmd}
\end{attackerbox}

The answer of the system was
\begin{defenderbox}
\begin{DefenderCmd}
AS14061 DigitalOcean, LLC
\end{DefenderCmd}
\end{defenderbox}

\section{Conclusions, Limitations, and Future Work}
\label{sec:conclusions}

This paper presented AdvancedShelLM, an LLM-based SSH honeypot that extends shelLM with a Worker-Manager architecture, multi-LLM support, explicit file system state, real-time awareness, and cross-session state preservation. These mechanisms substantially improved generative correctness outperforming the strongest original shelLM baseline.

However, the evaluation also shows that better command generation is not sufficient for deception. Human participants still noticed latency, uniform response timing, history and state inconsistencies, and occasional model artifacts. ARACNE was useful as an adaptive probing tool, but not as a ground-truth classifier, since it also misclassified the real Ubuntu baseline and was indecisive most of the time. The Internet deployment showed that AdvancedShelLM can sustain real SSH attacks without revealing its true nature, preserve state for returning sessions, and influence future behaviour of the attacks from the Internet.

The main limitations are latency, occasional state inconsistencies, limited sample size in the human evaluation, and the early state of AI-attacker evaluation. Future work could focus on reducing response time through caching, selective Manager bypass, or smaller supervisory models as well as researching new approaches to maintain command output consistency in longer conversations.

\section*{Ethical Considerations}
This evaluation was conducted with approval of the faculty ethics board. Claude and Codex were used for grammar checks during writing this manuscript. The authors take full responsibility for the content of the published article.

%
%
%
\bibliographystyle{splncs04}
\bibliography{references.bib,second-references}

\end{document}